\documentclass[12pt]{iopart}
\usepackage{graphicx}
\begin{document}

\title[]{ Study of Charge Dependent Azimuthal Correlations using Reaction-Plane-Dependent Balance Functions}

\author{Hui Wang for the STAR Collaboration }

\address{National Superconducting Cyclotron Laboratory, Michigan State University, East Lansing, MI, 48824-1321, USA}
\ead{wang@nscl.msu.edu}
\begin{abstract}

STAR has recently reported charge-dependent azimuthal correlations that are sensitive to the charge separation effect in Au+Au collisions at $\sqrt{s_{\rm NN}}$ = 200 GeV. Qualitatively, these results agree with some of the theoretical predictions for local parity violation in heavy-ion collisions. However, a study using reaction-plane-dependent balance functions shows an alternative origin of this signal. The balance function, which measures the correlation between oppositely charged pairs, is sensitive to the mechanisms of charge formation and the subsequent relative diffusion of the balancing charges. The reaction-plane-dependent balance function measurements can be related to STAR's charge-dependent azimuthal correlations. We report reaction-plane-dependent balance functions for Au+Au collisions at $\sqrt{s_{\rm NN}}$ = 200 , 62.4, 39, 11.5, and 7.7 GeV using the STAR detector. The model of Schlichting and Pratt incorporating local charge conservation and elliptic flow reproduces most of the three-particle azimuthal correlation results at 200 GeV. The experimental charge-dependent azimuthal charge correlations observed at 200 GeV can be explained in terms of local charge conservation and elliptic flow.

\end{abstract}

\section{Introduction}
Recently, it has been discussed that the hot and dense matter created in heavy ion collisions may form metastable domains where parity is locally violated.  This possible local parity violation \cite{LPV} coupled with a strong magnetic field produced by passing nuclei in such a collision could cause a charge separation across the reaction plane in non-central collisions called the chiral magnetic effect (CME) \cite{CME_1,CME_2,CME_3}. One observable proposed to measure the CME is the three point correlator $\gamma_{\alpha, \beta}=\left\langle\cos (\phi _\alpha  + \phi _\beta  - 2\psi _{RP} )\right\rangle $ \cite{3_point_correlator}.

On the other hand, the balance functions, which measure the correlation between the opposite sign charge pairs, are sensitive to the mechanisms of charge formation and the subsequent relative diffusion of the balancing charges \cite{balance_PRL}. The reaction-plane-dependent balance functions can be written as

\begin{eqnarray}
\fl
B(\phi ,\Delta \phi ) = \frac{1}{2}\{ \frac{{\Delta _{ +  - } (\phi ,\Delta \phi ) - \Delta _{ +  + } (\phi ,\Delta \phi )}}{{N_ +  (\phi )}} + \frac{{\Delta _{ -  + } (\phi ,\Delta \phi ) - \Delta _{ -  - } (\phi ,\Delta \phi )}}{{N_ -  (\phi )}}\} .
\end{eqnarray}

Here  $N_{+(-)}(\phi)$ is the total number of positive(negative) particles that have an azimuthal angle $\phi$ with respect to the event plane and $\Delta_{+-}(\phi,\Delta \phi)$ represents the total number of pairs summed over all events where the first (positive) particle has an azimuthal angle $\phi$ with respect to event plane and the second (negative) particle has a relative azimuthal angle $\Delta\phi$ with respect to the first particle. Similarly we can express $\Delta_{++}(\phi,\Delta \phi)$, $\Delta_{-+}(\phi,\Delta \phi)$ and $\Delta_{--}(\phi,\Delta \phi)$.

The data used in this analysis are from Au+Au collisions at $\sqrt{s_{\rm NN}}$ = 200 , 62.4, 39, 11.5, and 7.7 GeV taken by the STAR detector.  A transverse momentum cut of $0.2 < p_{t} < 2.0$ GeV/$c$ was applied as well as a pseudorapidity cut of $|\eta| < 1.0$. The second order event plane from TPC is used here and electrons are suppressed by specific energy loss inside the TPC.

\section{Results}

\begin{figure}[h]
\begin{minipage}[t]{18pc}
\begin{center}
\includegraphics[width=18pc]{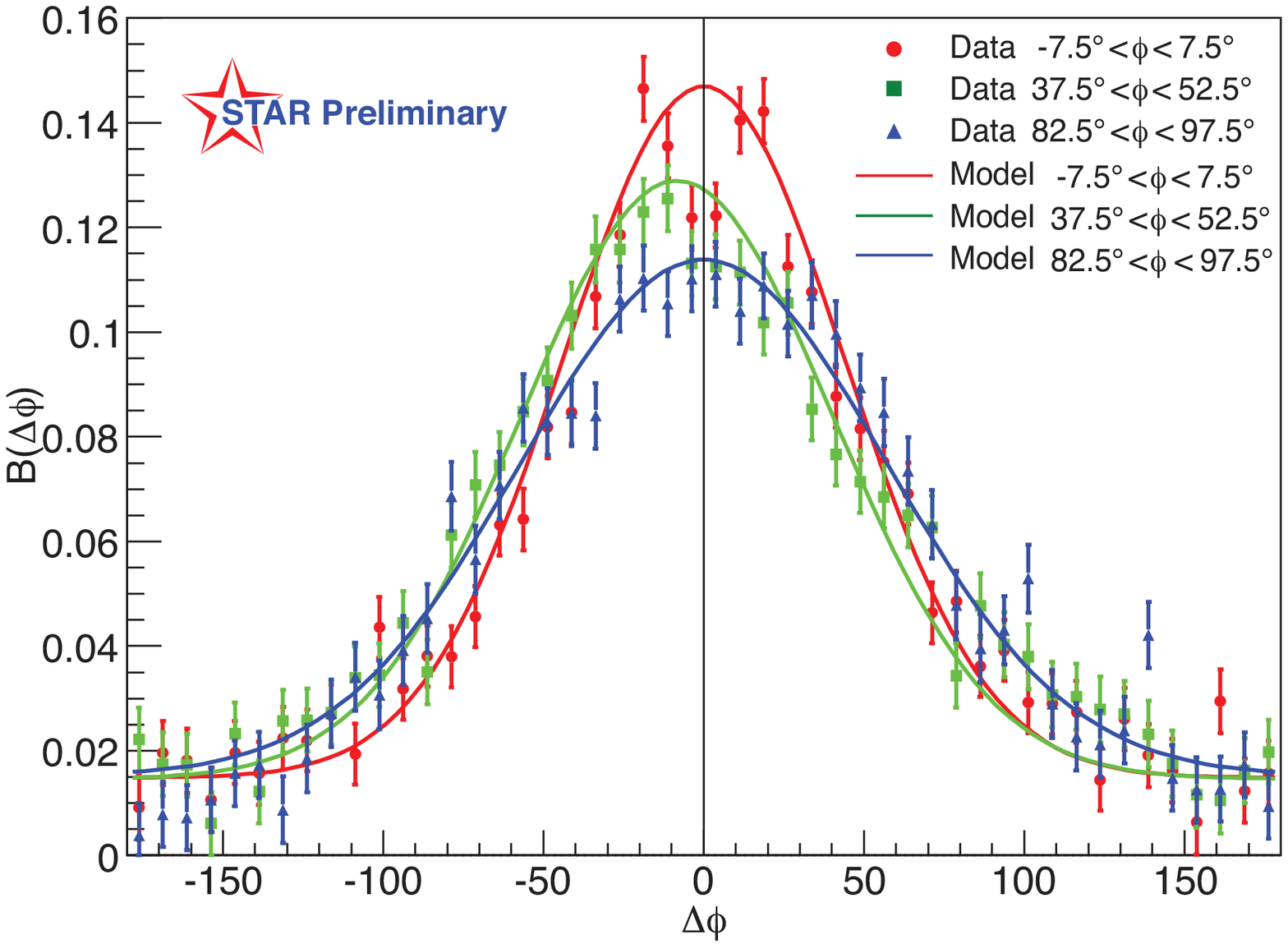}
\end{center}
\end{minipage}
\hspace{2pc}%
\begin{minipage}[t]{18pc}
\begin{center}
\includegraphics[width=18pc]{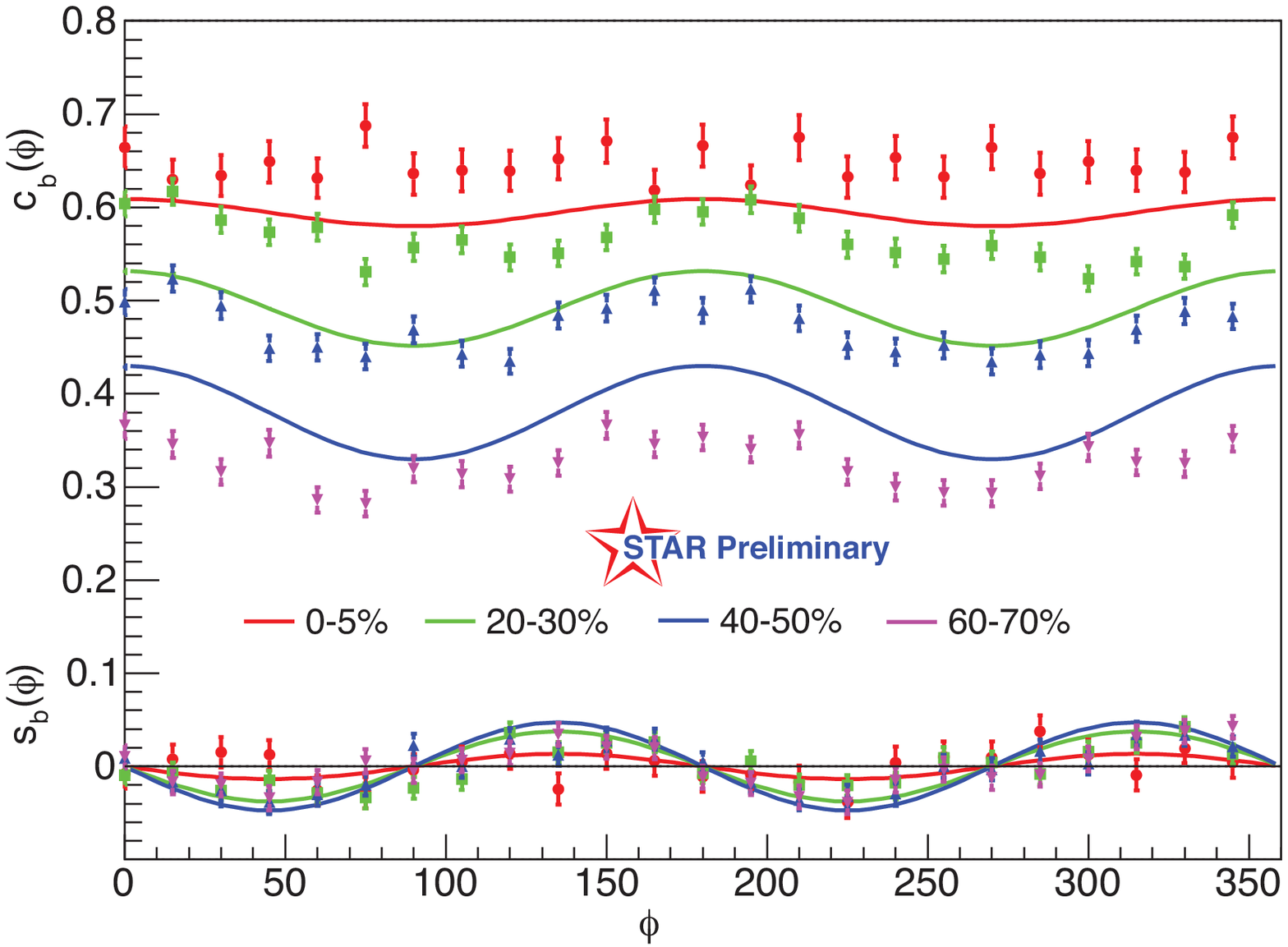}

\end{center}
\end{minipage} 
\caption{\label{fig:fig01}(Color online)  Left: The balance function for $\phi = 0^{\circ}$ (in-plane), $\phi = 45^{\circ}$, and $\phi = 90^{\circ}$ (out-of-plane) particles pairs (not corrected for event plane resolution). The 40-50\% centrality bin is shown. Right: The weighted average cosine and sine of balance function. Four centralities are shown here. The points are from the experimental data(not corrected for event plane resolution), while solid lines are from the blast-wave model of Ref. \cite{parity_soeren}.}
\end{figure}

The left panel of figure~\ref{fig:fig01} shows $\phi = 0^{\circ}$ (in-plane), $\phi = 45^{\circ}$, and $\phi = 90^{\circ}$ (out-of-plane) balance function for 40-50\% centrality only. The in-plane balance function is narrower than the out-of-plane balance function, which is caused by the strong collective flow in-plane, the $\phi = 45^{\circ}$ balance function is asymmetric and peaked at negative $\Delta \phi$ because charge pairs are more correlated on the in-plane side due to strong elliptic flow.  Also shown are the blast-wave model calculations from Ref. \cite{parity_soeren}.

To quantify the collective flow effect on balance function, we also study the weighted average cosine, $c_b(\phi)$, and sine, $s_b(\phi)$, extracted from the balance functions.
\begin{eqnarray}
\fl
\nonumber
c_b (\phi ) = \frac{1}{{z_b (\phi )}}\int {d\Delta \phi } B(\phi ,\Delta \phi )\cos (\Delta \phi ),\;\;\;s_b (\phi ) = \frac{1}{{z_b (\phi )}}\int {d\Delta \phi } B(\phi ,\Delta \phi )\sin (\Delta \phi ), \\
\fl
z_b (\phi ) = \int {d\Delta \phi B(\phi ,\Delta \phi )}.
\end{eqnarray}

$c_b(\phi)$ represents the width of balance function. If charges are created at the same point and do not diffuse due to strong collective flow, $c_b(\phi)$ would be close to unity.  $s_b(\phi)$ is an odd function of $\Delta\phi$, so it can quantify the asymmetry of the balance function. The right panel of figure~\ref{fig:fig01} shows $c_b(\phi)$ and $s_b(\phi)$ for Au+Au  collisions at  $\sqrt{s_{NN}}$ = 200 GeV.   $c_b(\phi)$ is closer to unity in the 0-5\% centrality bin, which is due to a stronger collective flow in central collisions, while in mid-peripheral and peripheral collisions,  $c_b(\phi)$ shows a difference between the in-plane and out-of-plane balance functions, which is caused by stronger elliptic flow in the event plane.  $s_b(\phi)$ reaches a maximum at $\phi {\rm{  =  }}135^{\circ},315^{\circ}$ and a minimum at $\phi {\rm{  =  }}45^{\circ},225^{\circ}$, which demonstrates that charged pairs are more likely to be emitted in-plane.

The right panel of figure~\ref{fig:fig01} also shows a comparison with the blast-wave model of Ref. \cite{parity_soeren}. The blast-wave model includes a breakup temperature $T_{kin}$, the maximum collective velocities in the in-plane and out-of-plane directions, the spatial anisotropy of the elliptic shape by fitting STAR published $v_{2}$ and spectra \cite{STAR_v2}.  This model also assumes local charge conservation and initial separation of balancing charges at freeze-out by fitting experimental results \cite{balance_PRC}. The difference between data and the blast-wave model predictions could be due to the finite event plane resolution for the data.

The difference between the same-sign and opposite-sign three point correlator $\gamma_{\alpha \beta}$ can be expressed as \cite{parity_soeren} 
\begin{eqnarray}
\fl
\gamma _p  = \frac{1}{2}(2\gamma _{ +  - }  - \gamma _{ +  + }  - \gamma _{ -  - } ) = \frac{2}{M}[v_2  < c_b (\phi ) >  + v_{2c}  - v_{2s} ],
\end{eqnarray}
\vspace{1pc}
where\;\;\;\; $ v_{2c}  =  < c_b (\phi )\cos (2\phi ) >  - v_2  < c_b (\phi ) >,  \;\;\;\;\;\; v_{2s} =  < s_b (\phi )\sin (2\phi ) >  $
and the bracket represents\;\;\;\;  $ < f(\phi ) >  = \frac{1}{M}\int {d\phi \frac{{dM}}{{d\phi }}z_b (\phi )f(\phi )}  $.

In this equation, $v_2 \left\langle {c_b (\phi )} \right\rangle $ will be positive if there are more charge pairs in-plane than out-of-plane, $v_{2c}$ will be positive if the charge pairs are more correlated in-plane than out-of-plane, while $v_{2s}$ will be negative if the charge pairs are more correlated on the in-plane side.

\begin{figure}[h]
\begin{minipage}[t]{18pc}
\includegraphics[width=18pc]{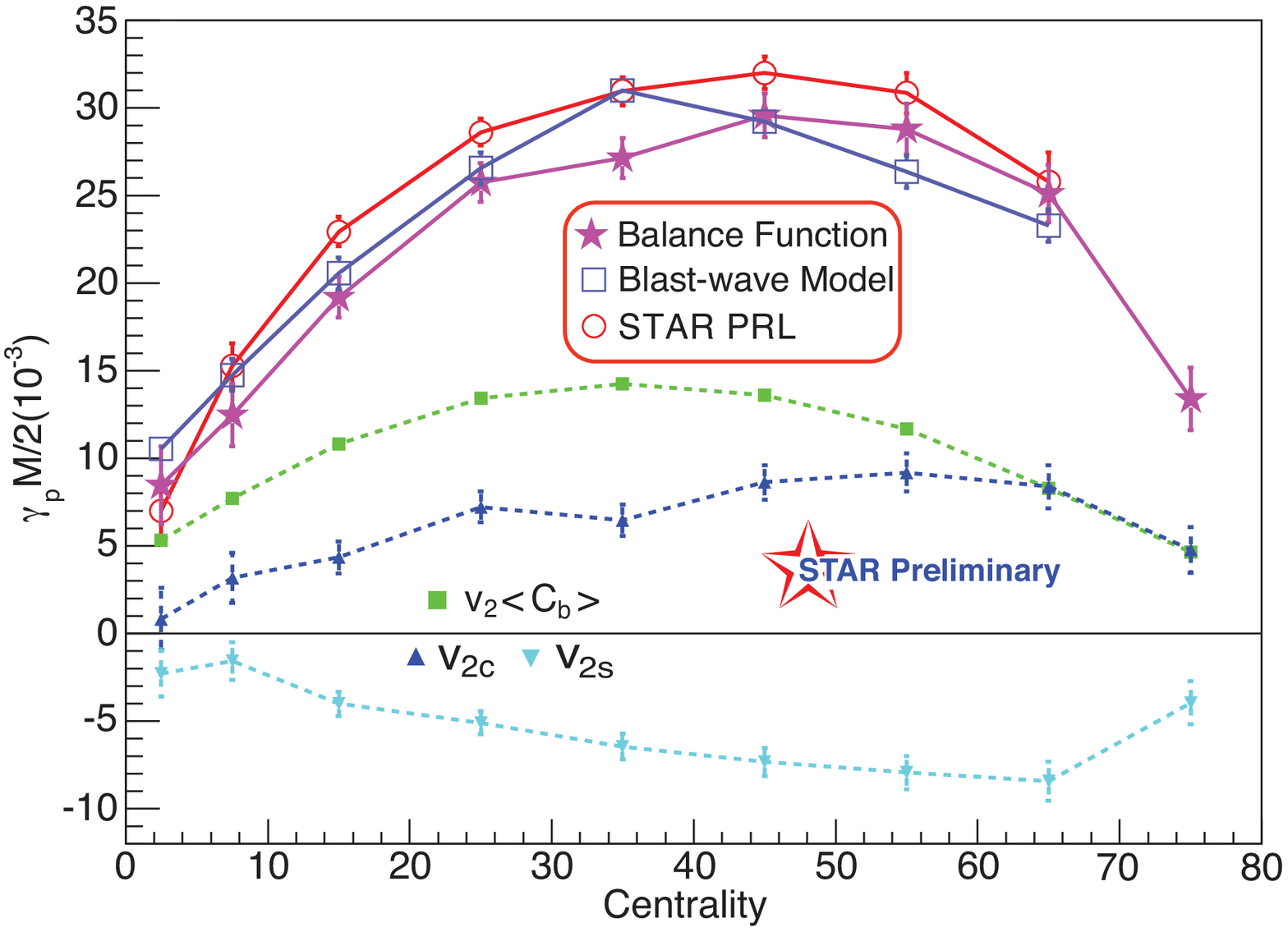}
\end{minipage}\hspace{2pc}%
\begin{minipage}[t]{18pc}
\includegraphics[width=18pc]{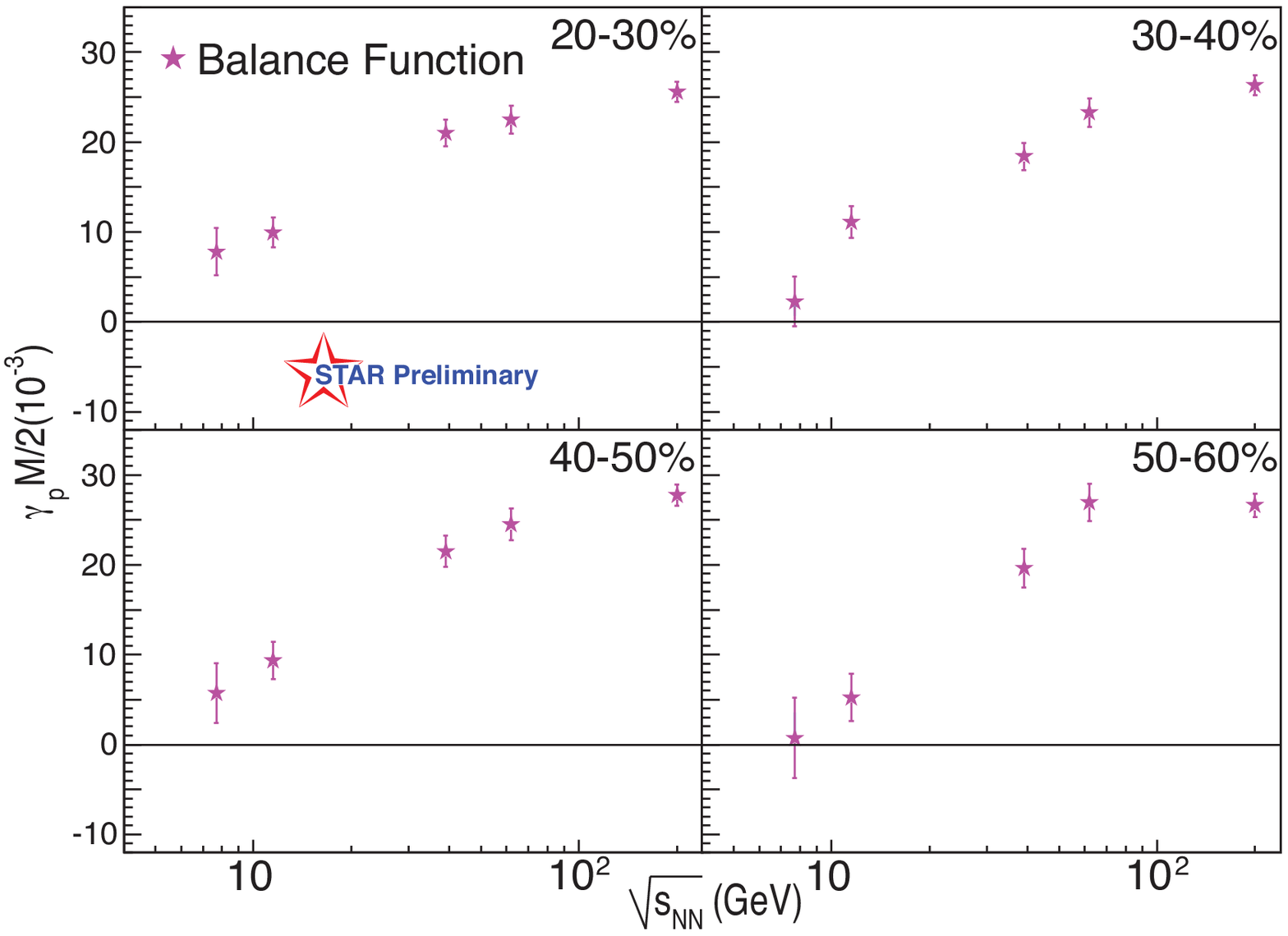}
\end{minipage} 

\caption{\label{fig:fig02}(Color online)  Left: Parity observable $\gamma_P$ scaled by experimental multiplicity. Right: The beam energy dependence of balance function.  Four centralities are shown here. }

\end{figure}

The left panel of figure~\ref{fig:fig02} shows the parity observable calculated from balance functions as well as its three components. All data points are corrected for the event-plane resolution here. To compare with previous results, we also plot the $\gamma_P$  from STAR published data \cite{parity_PRL} scaled by the measured uncorrected multiplicity in the same plot. Mathematically, the balance function result should equal the one from $\gamma_P$ and they do agree well. We can also see that a thermal blast-wave model \cite{parity_soeren} incorporating local charge conservation and flow reproduces most of the signal.

Another topic of interest is the beam energy dependence of the CME.  Recent calculations show it only exists in the deconfined, chirally symmetric phase \cite{CME_3} and decreases with increasing beam energy \cite{CME_BES}. The right panel of figure~\ref{fig:fig02} shows the same parity observable calculated from balance functions at $\sqrt{s_{\rm NN}}$ = 200 , 62.4, 39, 11.5, and 7.7 GeV. We can see that, for all four centralities shown here, the data show a smooth decrease with decreasing collision energy. This smooth decrease is different from the CME expectation which predicts an increasing signal with decreasing beam energy and sharply disappearance of signal near the top energy of SPS\cite{CME_BES}, but is consistent with the fact that $v_{2}$ decreases smoothly with beam energy in the same energy range.  

\section{Summary}
We have presented new results for the reaction-plane-dependent balance function. The reaction-plane-dependent balance function analysis gives the same difference between the like-sign and unlike-sign charge dependent azimuthal correlations as the three point correlator results published by STAR. A thermal blast-wave model incorporating local charge conservation and flow reproduces most of the difference between like- and unlike-sign charge-dependent azimuthal correlations. The reaction-plane-dependent balance function result shows a smooth decrease with decreasing collision energy.

\end{document}